\documentclass[conference]{IEEEtran}
\IEEEoverridecommandlockouts
\usepackage[utf8]{inputenc}
\usepackage{cite}
\usepackage{amsmath,amssymb,amsfonts}
\usepackage{algorithmic}
\usepackage{graphicx}
\usepackage{textcomp}
\usepackage{xcolor}
\usepackage{acro}
\usepackage{cleveref}
\usepackage{url}
\usepackage{csquotes}
\usepackage{tikz}
\usetikzlibrary{positioning,arrows,fit,backgrounds, shapes.geometric, calc}

\DeclareAcronym{LLM}{
    short=LLM,
    long=Large Language Model
}
\DeclareAcronym{DUT}{
    short=DUT,
    long=Device Under Test
}
\DeclareAcronym{IoT}{
    short=IoT,
    long=Internet of Things
}
\DeclareAcronym{SLT}{
    short=SLT,
    long=System-Level Test
}

\def\BibTeX{{\rm B\kern-.05em{\sc i\kern-.025em b}\kern-.08em
    T\kern-.1667em\lower.7ex\hbox{E}\kern-.125emX}}
\begin{document}

\title{
An Open-Source Power Measurement Platform for System-Level Semiconductor Testing
\thanks{This research was supported by Advantest as part of the Graduate School "Intelligent Methods for Test and Reliability" (GS-IMTR) at the University of Stuttgart.}
}

\author{\IEEEauthorblockN{Linus Bantel}
\IEEEauthorblockA{\textit{IPVS} \\
\textit{University of Stuttgart}\\
Stuttgart, Germany \\
linus.bantel@ipvs.uni-stuttgart.de}
\and
\IEEEauthorblockN{Sarah Rottacker}
\IEEEauthorblockA{\textit{Research and Venture}\\
\textit{Advantest Europe GmbH}\\
Stuttgart, Germany \\
sarah.rottacker@advantest.com}
\and
\IEEEauthorblockN{Dirk Pflüger}
\IEEEauthorblockA{\textit{IPVS} \\
\textit{University of Stuttgart}\\
Stuttgart, Germany \\
dirk.pflueger@ipvs.uni-stuttgart.de}
}

\maketitle

\begin{abstract}
Accurate power measurement is not only essential for evaluating the energy efficiency of modern embedded and semiconductor systems, but power draw is an important proxy during stress testing. 
Industrial semiconductor test equipment, however, is often expensive and difficult to integrate into flexible experimental workflows.

This paper presents a compact and extensible hardware platform for automated system-level power measurement of embedded devices.
The proposed setup integrates a Raspberry Pi controller, a precision current measurement device, and a microcontroller-based \ac{DUT}.
The system supports automated firmware deployment, synchronized device execution, and high-resolution current acquisition.

A lightweight HTTP-based interface allows remote clients to upload firmware binaries and retrieve measurement results.
This design enables automated benchmarking and reproducible experiments across multiple firmware variants.

The platform is implemented using accessible hardware components and open-source software tools, making it suitable for research, prototyping, and educational environments.
We describe the architecture of the platform, the automated testing workflow, and potential application scenarios, including firmware energy profiling and regression testing.
\end{abstract}


\section{Introduction}
Energy efficiency has become a central design objective in modern embedded systems and semiconductor devices. Numerous applications, such as \ac{IoT} nodes, mobile electronics, and edge-computing platforms, operate under strict energy budgets.
These systems frequently depend on battery power or energy harvesting, making careful management of energy consumption essential throughout both hardware and firmware development.

\Ac{SLT} executes realistic workloads and operational modes of the device in order to reveal failures that are difficult to detect with conventional structural test methods~\cite{Polian_2020}.
During \ac{SLT}, power consumption can serve as a practical indicator of the stress currently experienced by a chip.
When the objective is to deliberately provoke failure modes or expose reliability issues, monitoring the power draw can reveal abnormal operating conditions and provide insight into how the device behaves under stress.

In both contexts, accurate power measurement is therefore essential. Reliable measurements support optimization of firmware behavior and enable verification of power-management strategies.
Prior work has demonstrated that precise monitoring of supply current offers valuable insight into the runtime behavior of embedded devices~\cite{guo2021survey}, facilitating improvements in both hardware and software design.
Despite its importance, achieving precise current measurements in practice remains challenging. Conventional laboratory instruments, such as source measurement units (SMUs), oscilloscopes, and digital multimeters, can deliver high measurement accuracy, but they are not always well suited for all practical measurement scenarios.

These instruments are often expensive and difficult to integrate into automated testing pipelines.
Many measurement setups require manual interaction.
This makes them unsuitable for systematic benchmarking of multiple firmware configurations.
To overcome these limitations, prior research has explored low-cost and flexible measurement platforms for embedded devices.
These systems typically combine microcontrollers, measurement circuits, and data acquisition software to capture device power consumption~\cite{empiot,chia2018procal}.
While such platforms reduce cost and increase accessibility, many lack integrated support for automated firmware deployment.
They also do not provide coordinated execution control for reproducible measurements.
In this work, we present a compact hardware platform designed for automated power measurement of embedded devices.
The system integrates firmware flashing, device execution control, and high-resolution current acquisition into a unified workflow.
A Raspberry Pi controller coordinates communication between a precision CurrentRanger measurement device and the \ac{DUT}.
Firmware binaries can be uploaded remotely via an HTTP interface.
They are automatically flashed to the \ac{DUT}, executed, and analyzed through synchronized power measurements.
The platform enables fully automated and reproducible testing of multiple firmware versions.
The system provides the full time-current measurement allowing for indepth analytics and detailed metrics.
The main contributions of this work are:
\begin{itemize}
    \item Design of a compact and extensible hardware platform for system-level power measurement of embedded devices.
    \item Integration of automated firmware deployment with synchronized current measurement.
    \item Implementation of a remote testing interface enabling reproducible experiments.
    \item Demonstration of the platform for firmware energy profiling, benchmarking, and regression testing.
\end{itemize}
The remainder of the paper is structured as follows.
Section~\ref{sec:related} discusses related work in embedded energy measurement platforms.
Section~\ref{sec:architecture} presents the overall system architecture.
Section~\ref{sec:evaluation} presents results from a first study, showcasing the abilities of the proposed architecture.
Finally, Section~\ref{sec:limitations} discusses extensions for future projects.

\section{Related Work}
\label{sec:related}

Energy measurement and profiling of embedded systems has been an active research topic for many years.
Accurate measurement of device power consumption enables developers to optimize firmware implementations.
It also allows evaluation of energy-efficient algorithms and estimation of battery lifetime in embedded applications~\cite{guo2021survey}.
Several hardware platforms have been proposed for measuring the energy consumption of IoT and embedded devices.
For example, the EMPIOT platform provides a low-cost measurement system for wireless IoT devices.
It evaluates how hardware parameters, such as sampling rate and bus speed, affect measurement accuracy~\cite{empiot}.
Experimental results show measurement errors below 3.5\% compared to industrial reference equipment.
ProCal introduces a programmable calibration platform for IoT measurement devices.
It allows precise generation of current and voltage levels to calibrate energy monitoring systems~\cite{chia2018procal}.
This approach improves measurement accuracy while keeping hardware cost low.
Accurate current measurements are essential to characterize wireless communication protocols, sensor workloads, and sleep modes in battery-powered devices~\cite{block2026precise}.

Despite these advances, most existing systems focus primarily on measurement hardware.
They often lack integrated workflows for automated firmware deployment, coordinated execution, and synchronized data collection.
Our work addresses this gap by combining automated firmware flashing, remote test orchestration, and high-resolution current measurement within a unified experimental platform.
The platform enables reproducible and scalable testing of multiple firmware versions without manual intervention.
It also allows systematic energy profiling and benchmarking across embedded devices.

\section{System Architecture}
\label{sec:architecture}

\begin{figure}[t]
    \centering
    \includegraphics[width=0.99\linewidth]{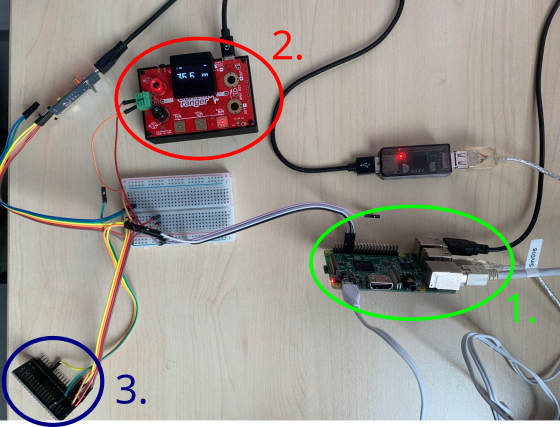}
    \caption{Prototype Setup: \textcolor{green}{1.} Raspberry Pi, \textcolor{red}{2.} CurrentRanger, \textcolor{blue}{3.} ESP32}
    \label{fig:rw_image}
\end{figure}

\begin{figure*}[t]
\centering
\begin{tikzpicture}[
node distance=3cm,
block/.style={draw, rectangle, rounded corners, minimum width=3cm, minimum height=1cm, align=center},
arrow/.style={->, thick},
dashedarrow/.style={->, thick, dashed}
]

\node[block] (client) {Client};
\node[block, right of=client, xshift=3cm] (rpi) {Controller \\ (Raspberry Pi)};
\node[block, below of=rpi, yshift=-1cm] (current) {Current Measurement \\ Device (CurrentRanger)};
\node[block, right of=rpi, xshift=3cm] (dut) {Device Under Test \\ (ESP32)};

\draw[arrow, <->] (client) -- node[align=center]{HTTP\\Commands} (rpi);
\draw[arrow,<->] (rpi.east) -- node[align=center]{Flash\\Serial} (dut.west);
\draw[arrow] (current.east) -|node[pos=0.75]{Power Supply} (dut.south);
\draw[arrow] (current.north) -- node{Current Samples} (rpi.south);

\draw[arrow] (rpi.north) -- ++(0,1cm)  -| node[above, pos=0.25]{Hardware Control Lines} (dut.north);

\end{tikzpicture}
\caption{Overview about the hardware and the connections between the modules.}
\label{fig:architecture}
\end{figure*}
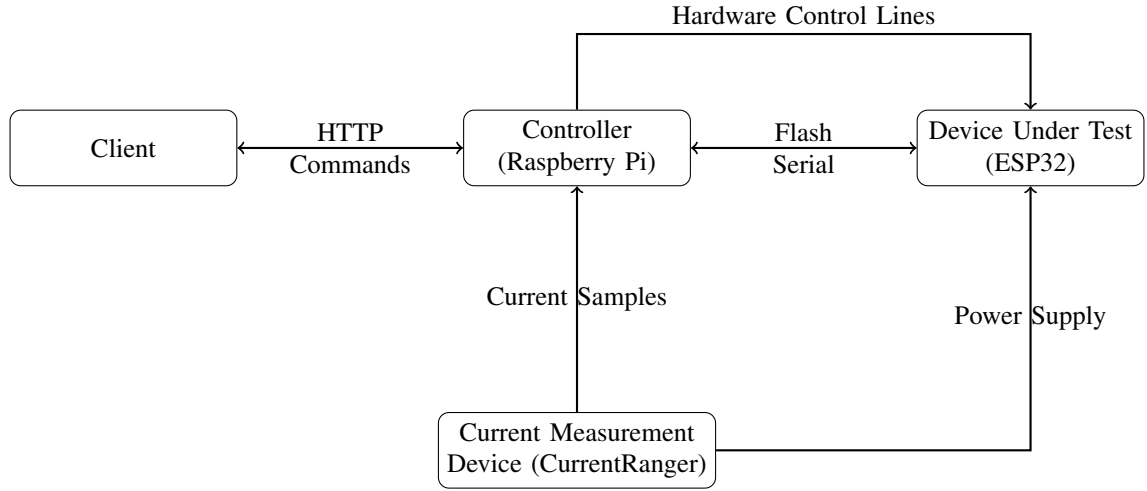

The proposed measurement platform is designed as a centralized test station.
It coordinates firmware deployment, device execution, and power measurement for embedded devices.
\Cref{fig:rw_image} shows the experimental setup.
\Cref{fig:architecture} illustrates the overall architecture of the system.
\subsection{Hardware Architecture}
The platform is composed of three main components:
\begin{itemize}
    \item Raspberry Pi 3~\cite{raspberrypi_website} acting as the system controller.
    \item CurrentRanger~\cite{lowpowerlab_currentranger_guide} as the measurement device.
    \item Embedded device under test (DUT), e.g., ESP32~\cite{espressif_esp32}.
\end{itemize}
The Raspberry Pi serves as the central coordination unit.
It hosts a lightweight control server implemented in Python.
The control server manages firmware upload, flashing, execution control, and data collection.
It provides a lightweight HTTP interface for external clients.
Clients can upload firmware binaries, trigger measurements, and retrieve results remotely.

The current measurement device is inserted directly into the DUT power supply path.
It continuously records the supply current with ca. 1200Hz during program execution.
This allows the platform to capture fine-grained current variations caused by computation, communication, and low-power sleep states.

The DUT is connected to the Raspberry Pi via a serial programming interface.
This interface is used both for firmware flashing and for optional debugging or logging.
The Raspberry Pi also controls the DUT's reset and boot mode pins through its GPIO interface.

This enables precise synchronization between the start of firmware execution and the start of current measurement.
Communication between the Raspberry Pi and the CurrentRanger occurs via USB or serial protocols.
The CurrentRanger streams timestamped current samples to the Raspberry Pi.
All measurement data is logged locally in CSV format.
The platform supports multiple measurement configurations, including adjustable sampling rate and measurement duration.

By combining serial communication, GPIO control, and automated data collection, the system achieves fully synchronized and reproducible testing.
This architecture allows researchers to evaluate multiple firmware versions or configurations without manual intervention.
The modular design also permits easy integration of additional sensors or future \ac{DUT} types.

\subsection{Software Architecture}
\label{sec:software}

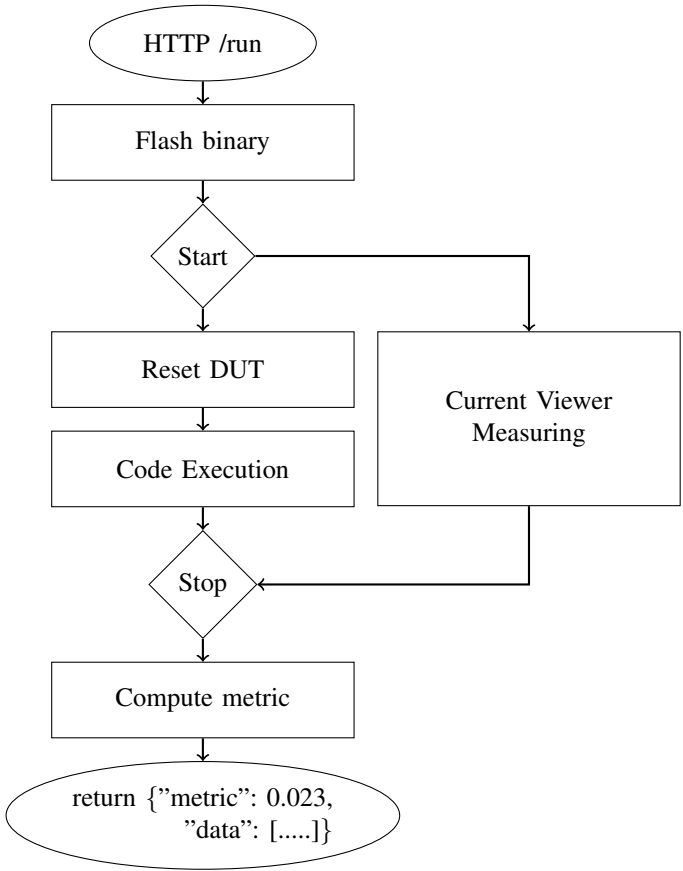
\begin{figure}[t]
\centering
\begin{tikzpicture}[
node distance=.3cm,
startstop/.style={ellipse, draw, minimum width=3cm, minimum height=1cm},
process/.style={rectangle, draw, minimum width=4cm, minimum height=1cm, align=center},
arrow/.style={->, thick}
]

\node (start) [startstop] {HTTP /run};
\node (flash) [process, below=of start] {Flash binary};
\node (measurestart) [draw, diamond, below=of flash] {Start};
\node (reset) [process, below=of measurestart] {Reset DUT};
\node (measure) [process, right=of reset, yshift=-0.65cm, minimum height=2.3cm] {Current Viewer\\Measuring};
\node (wait) [process, below=of reset] {Code Execution};
\node (stop) [draw, diamond, below=of wait] {Stop};
\node (return) [process, below=of stop] {Compute metric};
\node (end) [startstop, below=of return, align=right] {return \{"metric": 0.023,\\"data": [.....]\}};

\draw[arrow] (start) -- (flash);
\draw[arrow] (flash) -- (measurestart);
\draw[arrow] (measurestart) -- (reset);
\draw[arrow] (reset) -- (wait);
\draw[arrow] (wait) -- (stop);
\draw[arrow] (stop) -- (return);
\draw[arrow] (return) -- (end);
\draw[arrow] (measurestart) -| (measure);
\draw[arrow] (measure) |- (stop);

\end{tikzpicture}
\caption{Software architecture of the measurement control system.}
\label{fig:software}
\end{figure}

The measurement platform is controlled by a lightweight software stack running on the Raspberry Pi controller.
The software is responsible for orchestrating firmware deployment, device execution, current measurement, and result collection.
The overall design follows a modular architecture in which each subsystem is responsible for a specific stage of the testing workflow, see \Cref{fig:software}.   

At the core of the system is a Python-based control server implemented using the FastAPI framework.
This server exposes a simple HTTP interface that allows external clients to interact with the platform.
Through this interface, users can upload firmware binaries and initiate measurement runs remotely.
The server coordinates the subsequent steps required to deploy the firmware to the \ac{DUT}, execute the program, record current measurements, and return the resulting data to the client.

When a client submits a firmware binary, the control server stores the file temporarily and triggers the automated testing procedure.
In our case, firmware deployment is performed using the ESP-IDF toolchain, which provides utilities for flashing binaries to ESP32-based devices.
The toolchain for flashing can be easily replaced by other toolchains from different manufacturers.
The server invokes the flashing process programmatically, enabling the system to deploy firmware images without manual interaction.

In addition to firmware flashing, the Raspberry Pi also manages the hardware control signals required to operate the DUT.
The reset and boot configuration pins of the ESP32 are connected to the Raspberry Pi's GPIO interface.
A dedicated control module initializes these pins and provides functions for resetting the device and configuring the boot mode.
This mechanism enables the controller to synchronize the start of firmware execution with the beginning of current measurements.

Current measurements are recorded using the CurrentRanger and its open-source software CurrentViewer.
A measurement process continuously samples the current and stores the resulting data stream in a CSV log file.
This measurement process is executed as a separate subprocess to ensure that data acquisition proceeds independently of the control server.

After the firmware execution phase has completed, the recorded measurement trace is analyzed to extract relevant metrics.
A dedicated analysis module parses the CSV measurement data and computes statistics such as peak current consumption or other derived figures of merit.
These values provide a compact representation of the device's energy behavior during program execution.

The results of the measurement run are then returned to the client through the HTTP interface in JSON format.
This response typically includes both the computed metrics and optional measurement data for further analysis.

By integrating firmware deployment, device control, and power measurement into a unified software stack, the platform enables fully automated testing workflows.
This design allows researchers and developers to evaluate multiple firmware variants in a reproducible manner and facilitates systematic experimentation with embedded software energy consumption.

\section{Evaluation}
\label{sec:evaluation}

To demonstrate the capabilities of the proposed measurement setup, we generated multiple C code snippets using \acp{LLM} and executed the testing sequence via HTTP requests.
For each generated program, the firmware was automatically flashed to the device and the corresponding current trace was recorded.

\begin{figure}[t]
    \centering
    \includegraphics[width=0.99\linewidth]{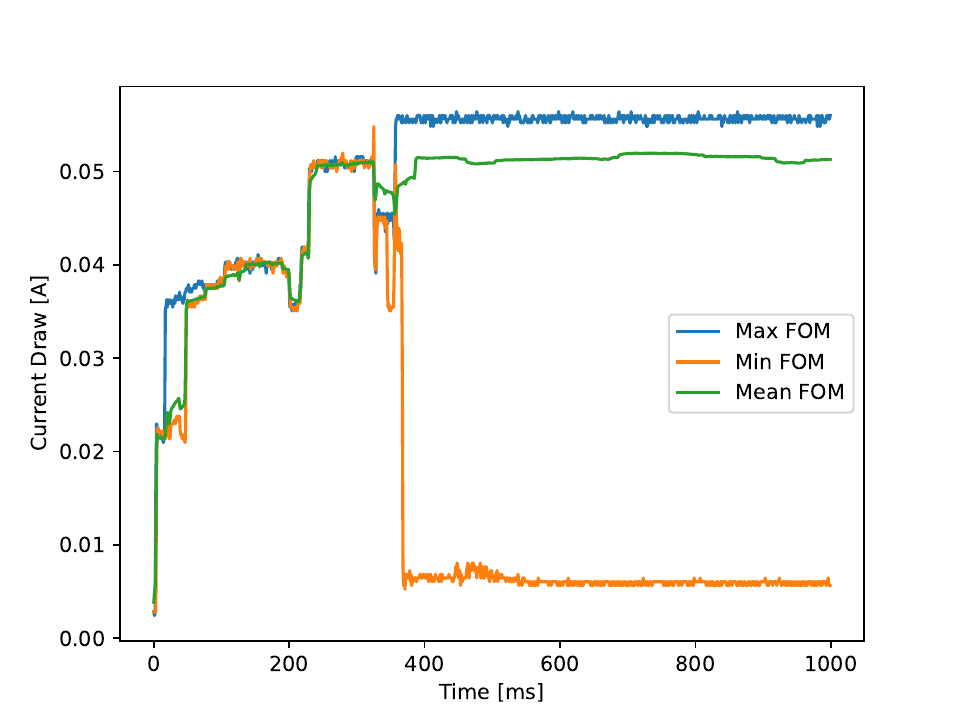}
    \caption{Current draw trace of the ESP32 for different code snippets. The trace with the highest/lowest average current consumtion, as well as the average trace are plotted. Until 350ms the start-up sequence of the chip can be seen, afterwards the current draw mainly depends on code snippet.}
    \label{fig:current_draw}
\end{figure}

The resulting measurements are shown in \Cref{fig:current_draw}.
Three representative curves are plotted: the trace with the lowest average current consumption, the trace with the highest average current consumption, and the average trace computed over all recorded measurements.

At the beginning of the traces, all three curves exhibit the same characteristic pattern.
This region corresponds to the start-up sequence of the ESP32 and is therefore independent of the executed code snippets.
After the initialization phase, the traces transition into a region with relatively stable current consumption that reflects the behavior of the individual programs.

A notable feature can be observed in the averaged trace, which exhibits a wavelike pattern after the startup sequence.
This effect is caused by several generated code snippets that lead to a crash of the ESP32, triggering an automatic restart.
As a result, the device repeatedly executes its start-up sequence, which becomes visible in the averaged current trace.

Overall, the results demonstrate that the automated workflow operates reliably.
The system successfully performs automated flashing and current sampling, while also correctly synchronizing the device reset with the measurement process.
This synchronization is clearly visible in the consistent start-up pattern at the beginning of the recorded traces.

\section{Limitations and Future Work}
\label{sec:limitations}

Despite the low cost, the current measurement setup still has lots of ways to enhance or extend the fidelity.

First, the system currently measures only the supply current rather than the actual power consumption of the device.
Since power is the product of voltage and current, variations in the supply voltage may introduce inaccuracies for high-power devices, when interpreting the measurements purely in terms of current draw.
Incorporating voltage measurements would allow direct calculation of power consumption and improve the reliability of the results.

Second, environmental parameters are not currently controlled.
Preliminary observations indicate that temperature changes can measurably influence the current draw of the device.
Without accounting for such external factors, it becomes more difficult to ensure reproducibility and comparability across measurements.
Future iterations of the setup should therefore incorporate environmental monitoring and, where possible, environmental control.

The set of observable system parameters can still be easily increased.
At present, only the current consumption is recorded.
Extending the setup with additional sensors—such as voltage monitoring, temperature sensing, and chip frequency measurements—would provide a more comprehensive view of the device's operating state.

Beyond passive monitoring, future work could also introduce more active control over experimental parameters.
For example, the ability to programmatically adjust supply voltage, ambient temperature, or processor frequency would enable more systematic exploration of the device’s behavior under different operating conditions.

Finally, the current synchronization between firmware execution and measurement is relatively coarse.
At the moment, synchronization relies on a single timestamp associated with the device start-up.
A more precise approach would involve explicit signaling between the firmware and the measurement system, for example via serial or GPIO lines.
Such mechanisms would enable instruction-level or event-based synchronization, allowing finer-grained analysis of how specific code segments influence power consumption.

\section{Conclusion}
\label{sec:conclusion}

We have presented a practical, cost-effective hardware platform for system-level semiconductor testing that integrates power measurement with automated code flashing and control.
The platform's open architecture allows for easy modification and extension, making it suitable for research and development environments.

The use of Raspberry Pi 3 as a controller provides sufficient computational resources for coordinating multiple tasks while remaining accessible to researchers and developers.
The CurrentRanger delivers accurate current measurements, and its integration with the existing ecosystem enables seamless workflow automation.

Future work will explore expansion to parallel testing scenarios and integration with advanced testing methodologies including machine learning-based anomaly detection in power signatures.

\bibliographystyle{ieeetr}
\bibliography{references}

\end{document}